\documentclass[amsmath,amssymb,aps,prl,reprint,superscriptaddress]{revtex4-2}

\usepackage{graphicx}
\usepackage{upgreek}
\usepackage{color}
\usepackage{setspace}
\usepackage{hyperref}
\hypersetup{colorlinks,breaklinks,
            urlcolor=[rgb]{0,0,0.64},
            linkcolor=[rgb]{0,0,0.64},
            citecolor=[rgb]{0,0,0.64}}
\usepackage{nccmath}
\usepackage{amssymb}
\usepackage{graphicx}
\usepackage{bm}
\usepackage{lineno}
\usepackage{ulem}
\usepackage{xcolor}

\usepackage{multirow}
\usepackage{makecell}

\makeatletter

\newcommand{\Rmnum}[1]{\expandafter\@slowromancap\romannumeral #1@}
\makeatother

\newcommand{\papertitle}{Magnetic switching of phonon angular momentum in a ferrimagnetic insulator}

\newcommand{\NJU}{National Laboratory of Solid State Microstructures and Department of Physics, Nanjing University, Nanjing 210093, China}
\newcommand{\IOP}{Institute of Physics, Chinese Academy of Sciences, Beijing 100190, China}
\newcommand{\UCAS}{University of Chinese Academy of Sciences, Beijing 100049, China}
\newcommand{\SLAB}{Songshan Lake Materials Laboratory, Dongguan 523808, China}
\newcommand{\COll}{Collaborative Innovation Center of Advanced Microstructures and Jiangsu Physical Science Research Center, Nanjing University, Nanjing 210093, China}

\begin{document}


\title{\papertitle}


\author{Fangliang~Wu}
\thanks{These authors contributed equally to this work.}
\affiliation{\NJU}

\author{Jing~Zhou}
\thanks{These authors contributed equally to this work.}
\affiliation{\IOP}
\affiliation{\UCAS}

\author{Song~Bao}
\affiliation{\NJU}
\affiliation{\COll}

\author{Liangyue~Li}
\affiliation{\NJU}

\author{Jinsheng~Wen}
\affiliation{\NJU}
\affiliation{\COll}

\author{Yuan~Wan}
\thanks{Correspondence to: Y.W. (\href{mailto:yuan.wan@iphy.ac.cn}{yuan.wan@iphy.ac.cn}).}
\affiliation{\IOP}
\affiliation{\UCAS}
\affiliation{\SLAB}

\author{Qi~Zhang}
\thanks{Correspondence to: Q.Z. (\href{mailto:zhangqi@nju.edu.cn}{zhangqi@nju.edu.cn}).}
\affiliation{\NJU}
\affiliation{\COll}

\date{\today}

\begin{abstract}
Circularly polarized phonons offer a new route for mediating angular momentum in solids. However, controlling phonon angular momentum without altering the material's structure or composition remains challenging. Here, we demonstrate the non-volatile switching of angular momentum-carrying phonons by leveraging intrinsic ferrimagnetism in an insulator. We find a pair of chiral phonons with giant energy splitting reaching 20\% of the phonon frequency, due to spontaneously broken time-reversal symmetry. With a moderate magnetic field, the phonon angular momentum of the two chiral phonon branches can be switched along with the magnetization. Notably, near the critical temperature, the effective phonon magnetic moment is enhanced, reaching 2.62~Bohr magneton, exceeding the moment of a magnon. A microscopic model based on phonon-magnon coupling accounts for the observations. Furthermore, we identify two types of phononic domains with opposite phonon Zeeman splitting and propose the existence of topologically protected phononic edge modes at domain boundaries. These results demonstrate effective manipulation of chiral phonons with magnetism, and pave the way for engineering chiral phononic domains on the micrometer scale.
\end{abstract}

\maketitle


Phonons can exhibit angular momentum through circular atomic motions\cite{zhang2014angular,zhang2015chiral}, and be either truly or falsely chiral \cite{barron2012cosmic}, depending on whether the angular momentum $J$ has a non-zero projection along the wavevector $k$. In either case, angular momentum-carrying phonons serve as a critical bridge connecting the spin and lattice degrees of freedom. They play a pivotal role in a variety of magnetic dynamics, including ultrafast demagnetization processes\cite{dornes2019ultrafast,tauchert2022polarized}, phonon-induced macroscopic magnetization\cite{basini2024terahertz,davies2024phononic,sasaki2021magnetization,luo2023large,juraschek2022giant,ren2021phonon,Juraschek2017Dynamical,Nova2017NatPhys}, and phonon-mediated exchange interactions\cite{kim2023chiral,jeong2022unconventional}, as well as in non-reciprocal phonon propagation\cite{xu2020nonreciprocal,chen2021propagating,Ohe2024Quartz} and heat transport\cite{zhang2015chiral,park2020phonon}. Therefore, controlling the characteristics of angular momentum-carrying phonons presents significant opportunities to engineer emergent phenomena in solids where the transport and conservation of angular momentum are important.

As charge-neutral and spinless bosons, phonons are inherently challenging to manipulate directly. This complexity is particularly pronounced for those truly chiral phonons\cite{ishito2023truly,ueda2023chiral}, which possess angular momentum and chirality linked by inversion symmetry intrinsic to their structural framework. While theoretical proposals suggest the potential for structurally engineered control of angular momentum-carrying phonons through methods such as interlayer sliding\cite{chen2023phonon}, strain\cite{rostami2022strain} and moir\'{e} engineering\cite{suri2021chiral}, practical demonstrations have yet to be realized. In contrast, phonons with oppositely directed angular momentum states connected by time-reversal symmetry (TRS) offer a viable pathway for control through magnetization or external magnetic fields. Such phonons can be either truly or falsely chiral, as discussed later. We call them chiral phonon hereinafter. Chiral phonons can exhibit large Zeeman effect via coupling to electrons\cite{Juraschek2017Dynamical,Anastassakis1972Morphic}. Experimentally, it has been observed in phonon modes with large magnetic moments\cite{cheng2020large,baydin2022magnetic,hernandez2023observation,Schaack_1976CeF3,schaack1977CeCl3}, as well as gate-tuned phonon polarization mediated by Landau levels\cite{sonntag2021electrical}. However, these field-induced chiral phonon splitting vanishes at zero magnetic field, due to modes with opposite phonon angular momentum relying on TRS breaking. Non-volatile control of phonon angular momentum without altering the structure or composition of the material remains elusive.

In this paper, we demonstrate the non-volatile magnetic switching of chiral phonons in a polar ferrimagnetic (FiM) insulator, Fe$_{1.75}$Zn$_{0.25}$Mo$_3$O$_8$ (FZMO). The control of chiral phonons is efficiently achieved by leveraging a large FiM Weiss molecular field through a moderate magnetic field. Specifically, through polarization-resolved magneto-Raman spectroscopy, we unveil a pair of chiral phonons with unexpectedly large energy splitting of 1.25 meV\cite{Bonini2023Frequency,Ren2024Adiabatic,che2024magnetic,Ning2024spontaneous}, which constitutes nearly 20\% of the phonon frequency. This splitting correlates with the system's magnetization and diminishes above the Néel temperature (T$_\text{N}$). Near T$_\text{N}$, the phonon magnetic moment (PMM) is enhanced, reaching 2.62 Bohr magnetons ($\mu_\text{B}$), surpassing the values of typical magnon or electron. A moderate magnetic field (2000~Oe) is sufficient to switch the phonon AM along with the magnetization. Combining the Raman spectroscopic mapping with magneto-optical imaging, we identify two types of phononic domains with opposite Zeeman shifts, correlated with FiM domains. We further provide a microscopic model to explain these findings, supplemented by Monte Carlo simulations that successfully reproduce the asymmetric Zeeman shifts of phonons. The possible edge state in magnetic and phononic domain walls is also discussed, based on the phonon band topology in the honeycomb lattice with TRS breaking.

\begin{figure}[htb]
	\includegraphics[width=0.48\textwidth]{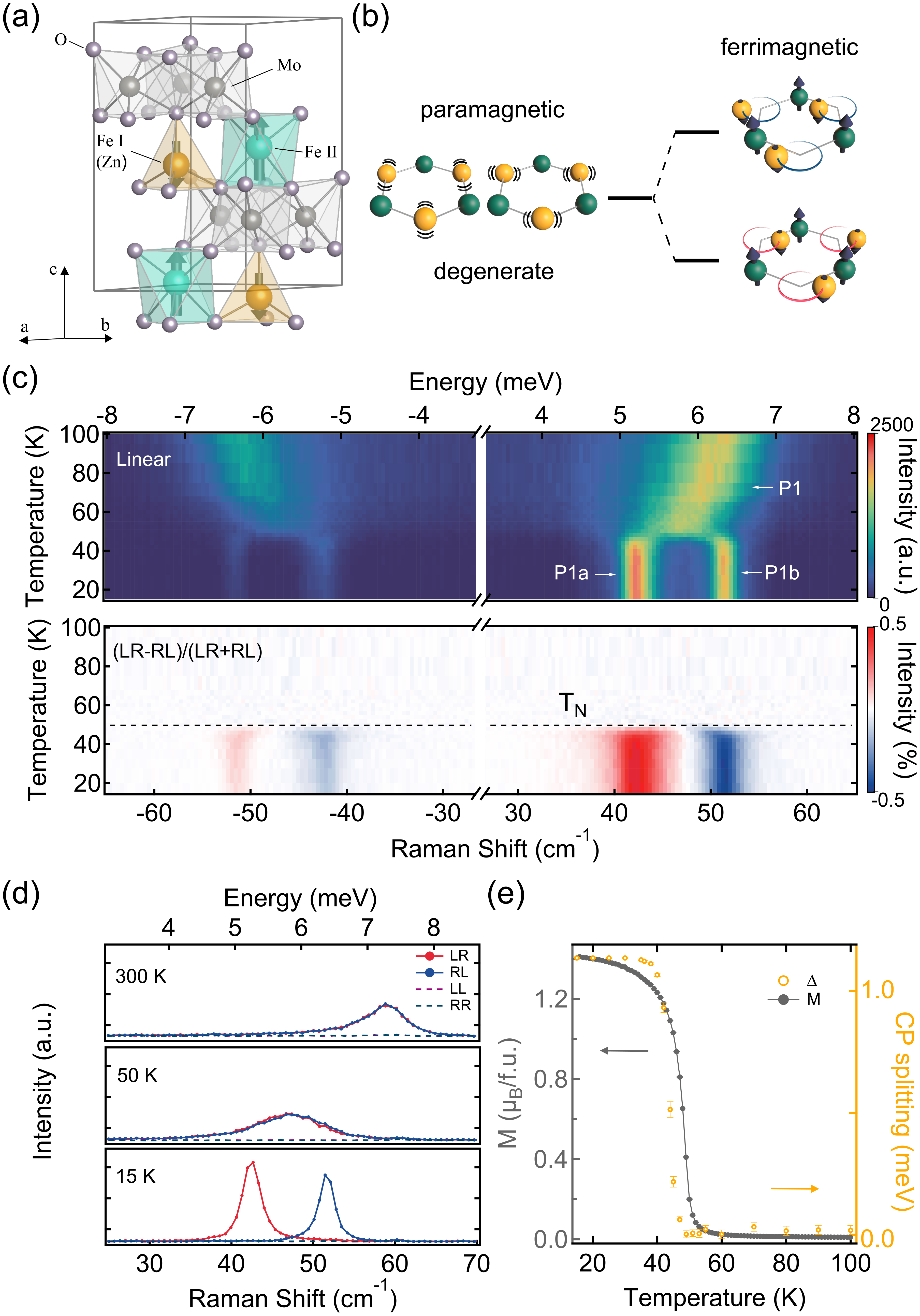}
	\caption{(a)~Crystal structure of Fe$_{1.75}$Zn$_{0.25}$Mo$_3$O$_8$ (FZMO). The black arrows indicate the magnetic moments of Fe ions. (b)~Schematic diagram for the P1 phonons. The yellow and green balls represent the Fe-\Rmnum{1} and Fe-\Rmnum{2}, respectively. (c)~Raman spectra as a function of temperatures. Upper panel: Raman intensity; Lower panel: degree of circular polarization of Raman spectra, defined as $(I_\text{LR}-I_\text{RL})/(I_\text{LR}+I_\text{RL})$. (d)~Circular polarization-resolved Raman spectra at 15~K, 50~K, and 300~K. (e),~The magnetization of FZMO under 0.1~T magnetic field (gray dots) and the P1 phonon splitting (yellow dots) as functions of temperatures.}
\label{fig:1}
\end{figure}

Ferrimagnetic FZMO is obtained through the Zn doping of its parent compound Fe$_2$Mo$_3$O$_8$, a polar antiferromagnetic insulator. This compound belongs to a pyroelectric space group \textit{P}6$_3$mc, characterized by electric polarization along the \textit{c}-axis\cite{Strobel1982JSSC}. As illustrated in Fig.~\ref{fig:1}{(a)}, the Fe-O tetrahedron(\Rmnum{1}) and Fe-O octahedron(\Rmnum{2}) form a honeycomb lattice structure. The  Mo-O octahedral layer separates the neighboring Fe layers, with the positions of \text{Fe-\Rmnum{1}} and \text{Fe-\Rmnum{2}} exchanged. Below 50~K, the N\'{e}el-type order hosted by Fe atoms dominates the magnetism of the system, with larger magnetic moments in octahedral sites. Preferential substitution of non-magnetic Zn atoms for tetrahedral sites gives rise to the ferromagnetic interlayer coupling, therefore, forming collinear FiM
 order along the \textit{c}-axis\cite{TokuraPRX2015,wang2015scirep}.

By performing circular polarization-resolved Raman spectroscopy on FZMO single crystals, we identified a pair of low-lying excitations at 42~$\text{cm}^{-1}$ (5.21~meV) and 51~$\text{cm}^{-1}$ (6.32~meV) remain almost unchanged in the FiM state but merge rapidly near T$_\text{N}$ (50K), as shown in Fig.~\ref{fig:1}{(c)}. For clarity, we will refer to the modes P1a and P1b (collectively P1)  hereinafter. The P1 mode retains a well-defined profile even at 300~K, suggesting its phonon origin, consistent with that of the parent compound Fe$_2$Mo$_3$O$_8$\cite{wu2023fluctuation,baosong2023}. Utilizing the cross-circular polarization configurations, specifically the left-handed excitation with right-handed detection (LR), and vice versa (RL), we map the Raman circular polarization using $(I_\text{LR}-I_\text{RL})/(I_\text{LR}+I_\text{RL})$, where $I_\text{LR}$ ($I_\text{RL}$) is the intensity in LR (RL) channel. The red and blue regions represent the polarization related to the LR and RL channel with opposite direction along the \textit{c}-axis. As shown in Fig.~\ref{fig:1}{(c)} lower panel, P1a and P1b exhibit opposite cross-circular Raman polarization, with the opposite selection rules in the Stokes and anti-Stokes processes, evidencing that P1a and P1b are a pair of angular momentum-carrying chiral phonons\cite{higuchi2011selection,wu2023fluctuation}. Moreover, P1 phonons can only be observed in the cross-circular polarization channels at all temperatures, consistent with their \textit{E}$_2$ symmetry (see Fig.~\ref{fig:1}{(d)}). In addition, P1 phonons exhibit an unusual blue shift in energy and linewidth broadening above T$_\text{N}$.

To explore the correlation between the magnetic order and chiral phonon splitting, we plot the energy difference between P1a and P1b as a function of temperature in Fig.~\ref{fig:1}{(e)} alongside magnetization. The data reveal that chiral phonon splitting emerges with the onset of FiM order. Notably, the breaking of TRS induces a large chiral phonon energy gap, reaching 20\% of the phonon frequency even in the absence of an external magnetic field. The interplay with magnetic order renormalizes the phonon energy and endows them with circular polarization, as schematically shown in Fig.~\ref{fig:1}{(b)}. The motion of two orthogonal phonons with \textit{E}$_2$ symmetry is mainly in the \textit{ab}-plane\cite{wu2023fluctuation}. In our experiment, the incident light is along the \textit{c}-axis, with the Raman detection performed in the backscattering geometry. Thus, despite the small wavevector of the incident photon ($k_{inc}$) compared with the size of the Brillouin zone, the observed P1 phonons possess a finite wavevector ($2k_{inc}$) along the \textit{c}-axis, aligning with the direction of the phonon angular momentum. Therefore, they are propagating chiral phonons along the high-symmetry path allowed by the three-fold rotational symmetry of the system\cite{chen2021propagating,ishito2023truly,ueda2023chiral}, namely, the truly chiral phonon. In contrast, the circularly polarized P1 phonons at $\Gamma$ point are falsely chiral\cite{barron2012cosmic}.

\begin{figure*}[htb]
	\includegraphics[width=0.95\textwidth]{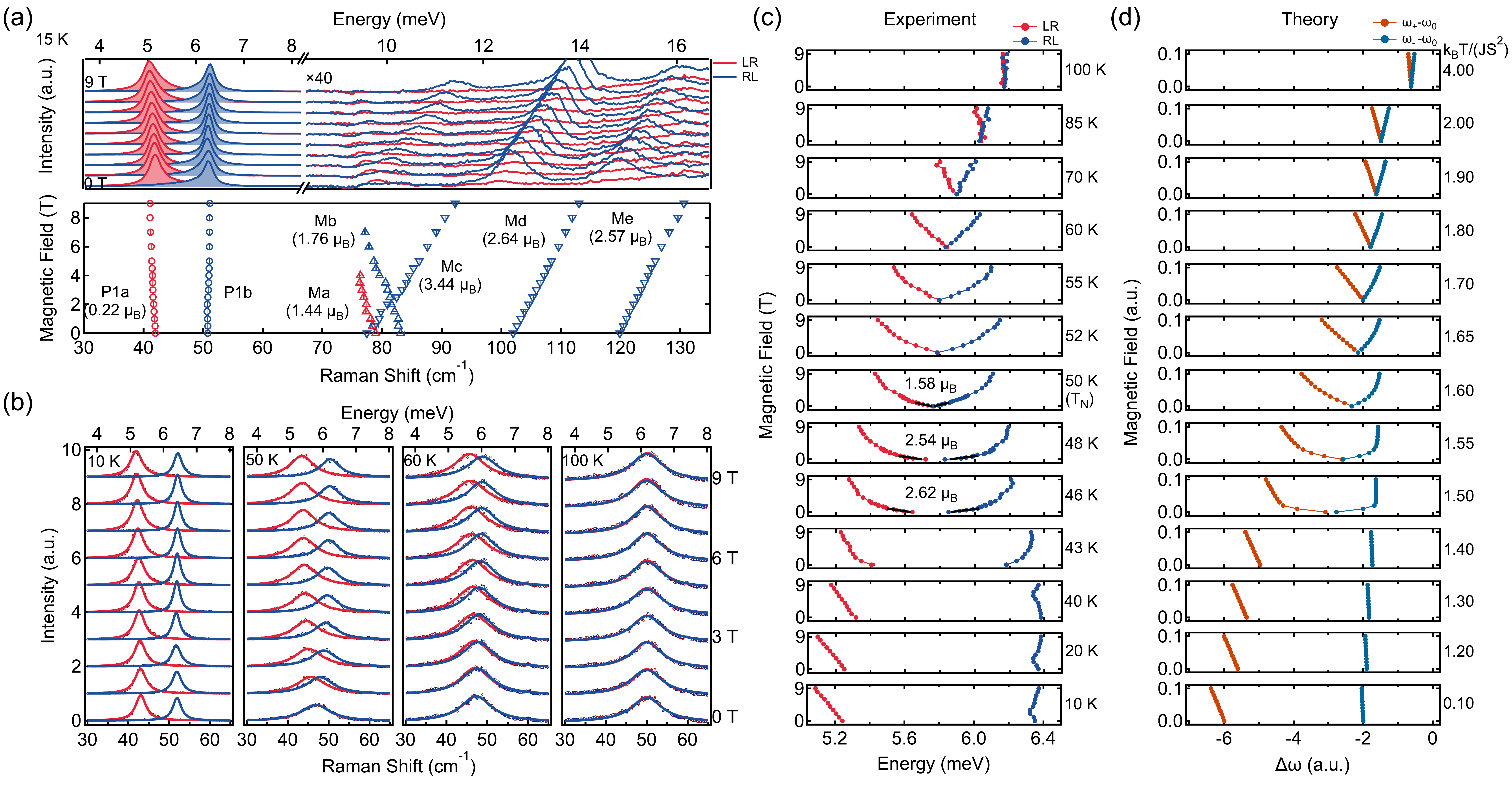}
	\caption{(a)~The magnetic field dependence of chiral phonons and magnons in Fe$_{1.75}$Zn$_{0.25}$Mo$_3$O$_8$ at 15~K. Fitted peak positions and effective magnetic moments are presented in the lower panel. Red (blue) curves/markers are from the LR (RL) channel. The error bars represent one standard deviation. (b)~Raman spectra of P1 phonon at representative temperatures. (c)~Peak position at different magnetic fields and temperatures. The phonon magnetic moments are extracted by taking numerical derivative of the phonon frequency with respect to the magnetic field near zero field, $\Delta$\textit{E}=$\mu_{ph}$${\cdot}$\textit{B}, where $\mu_{ph}$ is the phonon magnetic moment and \textit{B} is the increment of external. (d)~Simulated phonon frequency shift ($\omega-\omega_0$), where $\omega_0$ represents the bare phonon frequency.}
\label{fig:2}
\end{figure*}

The magnetic field dependence of the P1 phonon and nearby spin excitations further unveils the coupling between lattice and spin degrees of freedom in FZMO. As shown in Fig.~\ref{fig:2}(a), the two branches of P1 phonons exhibit asymmetric Zeeman shifts when an external magnetic field is applied along the \textit{c}-axis. P1a exhibits a pronounced linear shift, corresponding to a huge PMM of 0.22~$\mu_\text{B}$, which is three orders of magnitude larger than typical orbital PMM from ionic circular motion\cite{juraschek2019orbital}. In contrast, the P1b shows a smaller Zeeman slope. From 10 to 16~meV, five linear spin-wave branches, labeled as Ma to Me, are observed, similar in energy to those in the parent compound\cite{wu2023fluctuation,baosong2023}, and they disappear above T$_\text{N}$. In Fe$_2$Mo$_3$O$_8$, there are four spins per spin unit cell, corresponding to four linear spin-wave branches. The appearance of more than four magnon branches in FZMO likely results from multiple sublattices created by random Zn substitution at tetrahedral sites.

Figure~\ref{fig:2}(b) and (c) show the Raman spectra at representative temperatures and the extracted peak positions of P1 phonons, respectively. As the temperature increases and approaches T$_\text{N}$, the phonon energy shifts deviate from the linear Zeeman splitting, accompanied by a reduction in the chiral phonon gap. The splitting is enhanced at low fields, and saturated at high fields. Remarkably, a giant effective PMM of 2.62~$\mu_\text{B}$ is observed at low magnetic fields near T$_\text{N}$, which exceeds that of a one-magnon excitation. As the temperature rises further, the phonon Zeeman splitting restores its linearity and extends into the paramagnetic phase before disappearing at 100~K.

The observed chiral phonon splitting and asymmetric field dependence can be understood in terms of the interplay between chiral phonons and ferrimagnetism, driven by (1) phonon-magnon coupling and (2) FiM molecular field and spin fluctuation.

First, the pair of chiral phonons (P1) hybridizes with the nearby magnons, acquiring magnetic moments proportion to $(\gamma / \Delta)^2 \mu_\mathrm{mag}$, where $\gamma$, $\Delta$ and $\mu_\mathrm{mag}$ are the phonon-magnon coupling strength, the detuning frequency, and the magnon magnetic moment, respectively\cite{wu2023fluctuation,wang2024magnetic}. This coupling also explains the blue shift of P1 phonons above T$_\text{N}$, where it returns to its bare frequency without phonon-magnon interaction.

Second, the phonon-magnon coupling in this system is angular momentum selective, occurring primarily between phonon and magnon with the same angular momentum direction. This selectivity dictates P1a and P1b couple to different magnon branches, hosted by \text{Fe-\Rmnum{1}} and \text{Fe-\Rmnum{2}}. Assuming the coupling to Fe-I dominates, P1a exhibits field-induced red shift due to its hybridization with red shifting magnons on Fe-I. By contrast, at low temperature and low magnetic field, the blue shift of P1b is weaker than P1a due to mismatched angular momentum.

Third, the FiM ground state of FZMO generates a large molecular Weiss field, which directly causes the giant chiral phonon splitting at zero field. Near T$_\text{N}$, the system exhibits significant FiM fluctuation. An external magnetic field reinforces the FiM order, leading to the large phonon Zeeman shift at low fields and saturation at high fields. Therefore, the effective PMM is amplified by spin fluctuation.  Above T$_\text{N}$, the FiM spin fluctuation extends to the paramagnetic phase. This explains the Zeeman splitting of P1 phonons at high temperatures.

Figure~\ref{fig:2}{(d)} shows the Monte Carlo simulation of a minimal model based on this phonon-magnon coupling scheme. It well reproduces main observations including giant chiral phonon splitting, the asymmetric phonon Zeeman splitting, and the enhanced PMM near T$_\text{N}$. Detailed theoretical modeling and analysis can be found in the Supplementary Materials. We note the simulation does not capture the non-monotonic field-induced frequency shift of P1b phonon below T$_\mathrm{N}$, which we attribute to the complex spin-phonon couplings beyond this model.


\begin{figure}[htb]
	\includegraphics[width=0.48\textwidth]{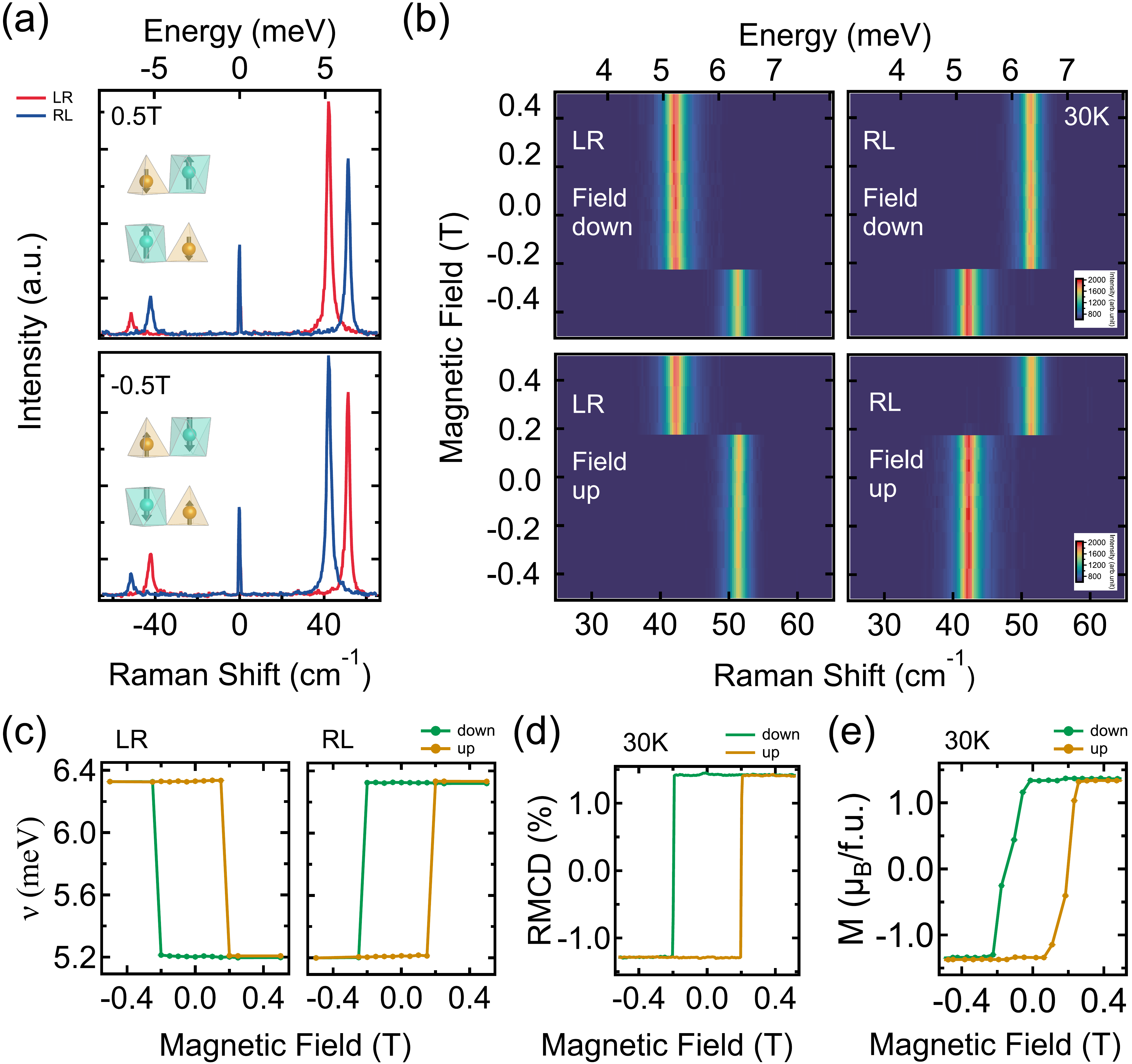}
	\caption{Magnetic control of chiral phonons. (a)~Raman spectra of P1 phonons with opposite ferrimagnetic polarization at 30~K. (b)~The chiral phonon hysteresis with an external magnetic field. (c)~The chiral phonon hysteresis loops of which peak position is fitted from raw data in (b) with Lorentz function. (d) and (e) Independent measurements of the reflected magnetic circular dichroism (RMCD) and magnetization, respectively. The error bars represent one standard deviation and are smaller than the symbols.}
\label{fig:3}
\end{figure}

The chiral phonon splitting in FZMO originates from TRS breaking and inherently depends on both spin correlation and fluctuation. It provides new routes for manipulating the property of chiral phonons via magnetic means. As shown in Fig.~\ref{fig:3}{(a)}, when the FiM order is reversed, the activated modes in LR and RL channels are switched accordingly. In Raman scattering, the pseudo-angular momentum (PAM) is conserved in an Umklapp-like process\cite{higuchi2011selection,tatsumi2018conservation,wu2023fluctuation}. The \textit{C}$_\text{3}$ symmetry defines chiral phonons with $\pm\hbar$ PAM. Hence, the switched activated modes in LR and RL channels indicate that the PAM of P1a (P1b) is switched from $+\hbar$($-\hbar$) to $-\hbar$ ($+\hbar$). At the long wavelength limit, the phonon PAM has a definite correspondence with phonon angular momentum, which is defined based on atomic circular motion\cite{zhang2014angular,zhang2015chiral}. Thus, we achieve non-volatile control of chiral phonon energy and angular momentum. The hysteresis of P1 phonons is observed by sweeping the magnetic field. As shown in Fig.~\ref{fig:3}{(b)}, for the LR channel, the P1a mode initially appears at 42~$\text{cm}^{-1}$ is switched to 51~$\text{cm}^{-1}$ at -0.2~T, while a conjugate scenario occurs in the RL channel for P1b phonon. The chiral phonon hysteresis is synchronized with the FiM hysteresis, which is independently verified by the reflected magnetic circular dichroism (RMCD) and magnetization measurements, as demonstrated in Fig.~\ref{fig:3}{(c-e)}.

\begin{figure}[htb]
	\includegraphics[width=0.48\textwidth]{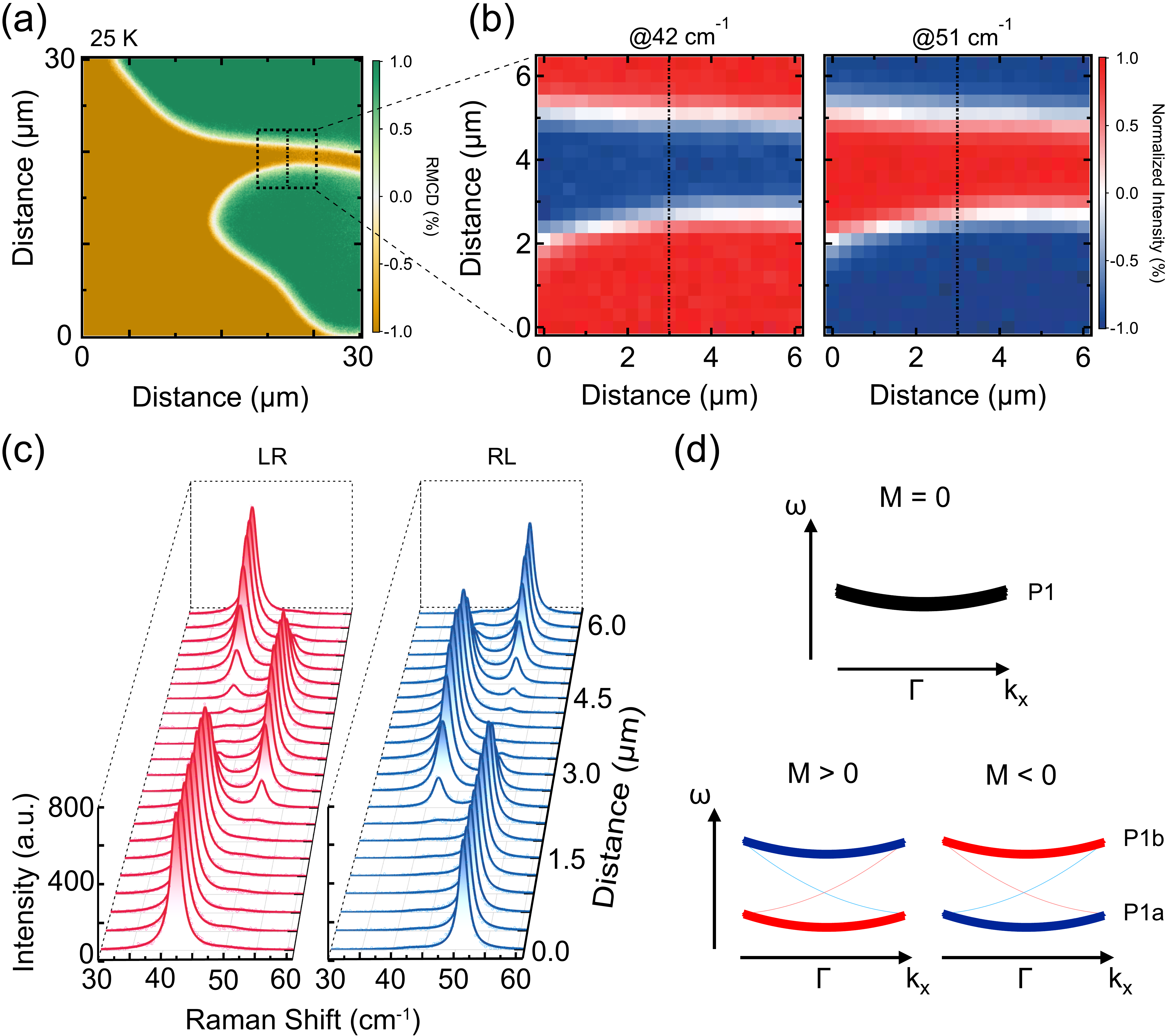}
	\caption{Observation of chiral phononic domains. (a)~Magneto-optical imaging of FiM domains in FZMO at 25~K. The green and yellow regions indicate the two types of domains with out-of-plane magnetization. (b)~The circular polarization mapping of Raman modes at 42~$\text{cm}^{-1}$ and 51~$\text{cm}^{-1}$, corresponding to the dashed area in FiM domain mapping. (c)~Line-scan of P1 phonon Raman spectra across FiM domains. The scanning line is indicated by dashed lines in (a) and (b). (d)~Schematic diagram for possible phononic edge state around $\Gamma$ point, denoted lines inside the bulk chiral phonon gap.}
\label{fig:4}
\end{figure}

So far, we have demonstrated the non-volatile manipulation of chiral phonon splitting with a moderate magnetic field (0.2~T). It promotes further exploration of spatial modulation of chiral phonons with FiM domains. After field-cooling to 10~K, the stripy-like FiM domains gradually appear upon warming to 25~K. As shown in Fig.~\ref{fig:4}{(a)}, the yellow and green regions represent the FiM domains with opposite orientations. Focusing on a smaller region containing up-down-up domains, we reveal two types of chiral phonon domains that are locked with the FiM domains, as visualized by Raman mapping shown in Fig.~\ref{fig:4}{(b)}. Figure~\ref{fig:4}{(c)} shows the Raman spectra from the scanning line (dashed), a significant shift in spectral weights is observed when the scan crosses the domain boundary. These findings pave the way for engineering chiral phononic domains on the micrometer scale by controlling the FiM domains, and further regulating the transport of chiral phonons and the flow of angular momentum.

Lastly, besides the opportunities enabled by chiral phononic domain engineering, these domains may possess non-trivial topological properties. The sign of chiral phonon gap switches between the FiM domains, which is reminiscent of the band inversion in a topological insulator. Here, we discuss the possibility of the existence of phononic edge mode at the domain boundary in FZMO. For a two-dimensional honeycomb lattice with TRS breaking, two-fold band degeneracy at $\Gamma$ point could be lifted, and generate topologically nontrivial phonon bands\cite{liu2017model,xiong2022interband}. We propose that FZMO can be a candidate system for implementing such a model (supplement provides the details). As schematically shown in Fig.~\ref{fig:4}{(d)}, the TRS breaking lifts the two-fold band degeneracy of P1 phonons, and edge states could emerge in the band gap. Once the magnetization is reversed, the edge state propagation direction is switched accordingly. Experimental identification of such edge mode can be challenging. The width of the FiM domain boundary is far below the diffraction limit of far-field Raman spectroscopy, which explains the absence of in-gap states in the line-scan spectra across chiral phononic domains. We envision the tip-enhanced Raman technique may resolve the in-gap edge modes at the domain boundary.

\begin{acknowledgments}
This work was supported by the National Key Research and Development Program of China (Grant Nos.~2020YFA0309200, 2022YFA1403800, 2021YFA1400400, and 2024YFA1409200), the National Natural Science Foundation of China (Grant Nos.~12474475, 12250008, 12434005, 12225407, and 12404173), the Natural Science Foundation of Jiangsu Province (Nos. BK20240057, BK20243011, BK20233001, and BK20241251) and the Fundamental Research Funds for the Central Universities.
\end{acknowledgments}

\normalem

\end{document}